# DESIGN OF AN INTERACTION REGION WITH HEAD-ON COLLISIONS FOR THE ILC*

**R. Appleby**
*Cockcroft Institute and the University of Manchester, Oxford Road, Manchester M139PL, England*

**D. Angal-Kalinin, F. Jackson**
*CCLRC, ASTeC, Cockcroft Institute, Daresbury Laboratory,
Chilton, DIDCOT, OX11 0QX, England*

**M. Alabau-Pons, P. Bambade, J. Brossard, O. Dadoun, C. Rimbault**
*Laboratoire de l'Accélérateur Linéaire,
IN2P3-CNRS et Université de Paris-Sud 11- Bât. 200, BP 34, 91898 Orsay cedex, France*

**L. Keller, Y. Nosochkov, A. Seryi**
*Stanford Linear Accelerator Center, 2575 Sand Hill Road, Menlo Park, CA 94025, USA*

**J. Payet[#], O. Napoly, C. Rippon, D. Uriot**
*CEA/DSM/DAPNIA-Saclay, 91191 Gif-sur-Yvette, France*

*Abstract*

An interaction region (IR) with head-on collisions is considered as an alternative to the baseline configuration of the International Linear Collider (ILC) which includes two IRs with finite crossing-angles (2 and 20 mrad). Although more challenging for the beam extraction, the head-on scheme is favoured by the experiments because it allows a more convenient detector configuration, particularly in the forward region. The optics of the head-on extraction is revisited by separating the e+ and e- beams horizontally, first by electrostatic separators operated at their LEP nominal field and then using a defocusing quadrupole of the final focus beam line. In this way the septum magnet is protected from the beamstrahlung power. Newly optimized final focus and extraction optics are presented, including a first look at post-collision diagnostics. The influence of parasitic collisions is shown to lead to a region of stable collision parameters. Disrupted beam and beamstrahlung photon losses are calculated along the extraction elements.

## INTRODUCTION

From an accelerator point of view, the main advantage of colliding the beams head-on rather than with a large crossing-angle, is to avoid relying on the RF "crab-crossing" technique to recover most of the luminosity, and to have the beam and detector solenoid axes aligned which prevents the beam from being deflected by the solenoid. The deflection of the beam-beam secondaries also compromises the very forward hermeticity of the detector in the case of a large crossing-angle. Hence a head-on collision scheme can in principle achieve a better reach for physics channels requiring detector coverage at very low polar angle. On the other hand, IR layouts with head-on collisions imply that the extraction line shares magnets with the final focus system over a considerable distance until beam separation is achieved and magnets allowing independent control of the extracted beam can be inserted. The low energy spent beam tail induces unavoidable losses from the first dispersive extraction element on. For specific collider optimisations [1] with enhanced beam-beam effects and increased beamstrahlung, beam losses can thus be a limiting factor.

The head-on collision geometry presented here (Fig. 1) is based on deflecting the outgoing beams horizontally with electrostatic separators similar to those used at LEP, located behind the final doublet and followed by a defocusing quadrupole centred on the incoming beam line to produce a complementary deflection before extraction in a septum magnet. In this scheme, beam losses occur far away from the IP and the septum magnet can be protected from the cone of beamstrahlung photons, an important improvement over the TESLA TDR design [2]. With a reduced 25 kV/cm electrostatic field, the beam separation at the first parasitic crossing is sufficient as long as the ILC train 308 ns bunch spacing is not much reduced. The performance of the extraction scheme is studied for 250 GeV beam energy. Its operation at 500 GeV will require upgrading the final doublet and the electrostatic separators.

*Work supported by the EC under the FP6 "Research Infrastructure Action - Structuring the European Research Area" EUROTeV DS Project Contract no.011899 RIDS
[#]jpayet@cea.fr

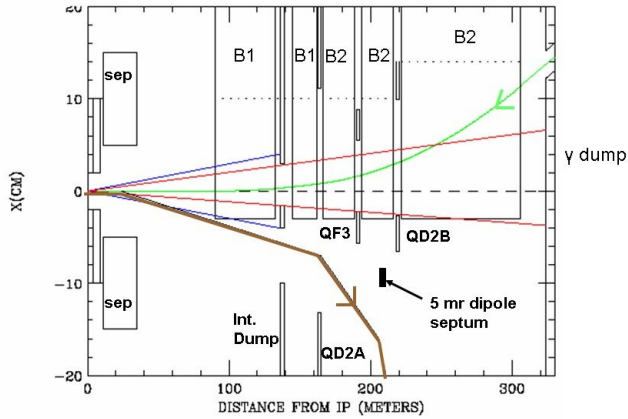

*Figure 1: Plan view of zero degree extraction from IP to beamstrahlung dump.*

## FINAL FOCUS OPTICS

The final focus system is adapted from the 2 mrad beam transfer line [3] to match the head-on requirements. The optimisation concerns the final doublet (FD) region and the chromatic correction. Choosing $L^*$ to be 4 m and the FD magnets to be superconducting (SC) quadrupoles with 56 mm bore diameter and 250 T/m gradient [4], their separation is set to 1.2 m to keep rather short magnets (1.24 and 0.72 m lengths) and provide enough space for a pair of SC sextupoles, correctors and instrumentation. The collimation depths are set to $14\sigma_x$ times $62\sigma_y$ by the 24 mm diameter e+e- pair masks inboard of the doublet. The 25 m long electrostatic separator is positioned 2 m behind the FD package in order to accommodate the kickers needed for the IP fast feedback system.

The linear optics of the beam delivery transfer line is shown on Fig. 2. Six sextupoles are used to correct the chromaticity and compensate the aberrations: two SC sextupoles with 3T pole tip field at 28 mm radius [5] are inserted in the final doublet, three others at the high beta upstream points, and one in the middle of the energy collimation section to improve the overall correction.

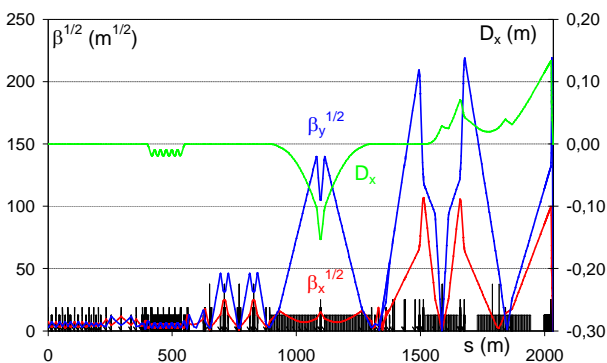

*Figure 2: Linear optics of the collimation and final focus beam line.*

The optimization of luminosity calculated from a particle cloud transport was done with the code TRACEWIN [6], by a standard minimization method with the beam line magnetic elements as variables. The resulting bandwidths are shown by Fig. 3 for several ILC parameter sets.

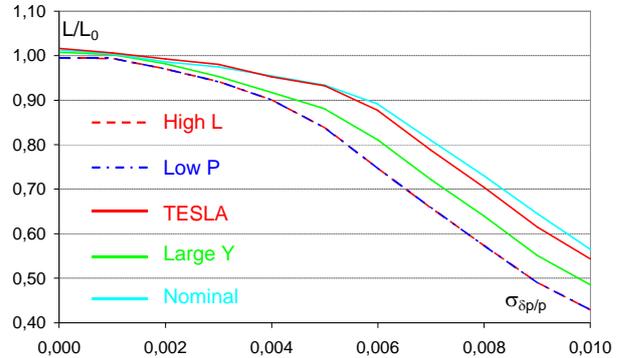

*Figure 3: Normalized luminosity as function of the beam RMS momentum spread for several ILC parameter sets.*

## ELECTROSTATIC SEPARATOR

The first extraction stage uses electrostatic separators with ±125 kV across a 10 cm gap, following LEP experience [7]. There would be five 5 m long modules, each enclosed in a 0.0166 T dipole which provides half of the total 0.5 mrad bend in the outgoing beam and cancels the bend of the incoming beam.

LEP was operated for many years with forty 4 m long separator modules. The required pressure in the separators is lower than $10^{-9}$ mbar (LEP had $10^{-10}$ mbar). At 30 kV/cm operating field, the breakdown rate was less than 0.01/hr for 3 mA, 100 GeV beams. The separators operated successfully in a flux of synchrotron radiation which drew several hundred µA from the high voltage power supply. It was estimated that $10^{17}$/s synchrotron photons with critical energy of 70 keV hit the plates [8]. In our case there is a 2.0 cm full-width, horizontal-gap synchrotron radiation mask near the inboard end of the separator which keeps outgoing synchrotron radiation originating in the final doublet from directly impacting the separator plates. This mask also blocks synchrotron radiation from the incoming beam from hitting the IP beam pipe. Simulations of the current drain caused by charged particles directly impacting the separator plates are ongoing. In practice the CERN group limited the maximum voltage to about ±220 kV. With higher voltages significant problems with high voltage cables, feed-through supports, geometrical field enhancement arise, and the breakdown rate at 50 kV/cm was about 0.2/hr with no beam.

With 25 kV/cm electrostatic field, this first stage provides 11 mm transverse separation at the first parasitic crossing 46 m from the IP. The effect of parasitic crossings coupled to the IP kink instability on the stability of the collision has been calculated [9] for increasing vertical beam jitter in the ILC train, showing a small additional luminosity loss (Fig. 4). A study done for a fixed offset confirms that 8 mm is the minimum necessary separation to stabilize the collisions.

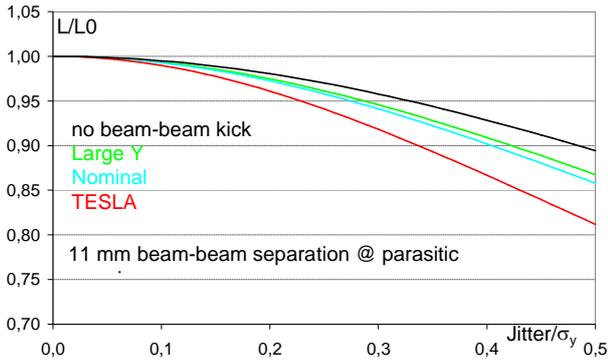

*Figure 4: Normalized luminosity as function of the vertical beam jitter, for several ILC beam parameters.*

## EXTRACTION OPTICS

After the first bend from the electrostatic separators, the outgoing beam passes to the outside of the final focus soft bend magnets, and off-axis through the horizontally defocusing quadrupole QD2A (Fig. 1). This gives the outgoing beam the second part of the beam-separating bend with an outward kick of 1.7 mrad. A dipole septum magnet of 5 mrad total bend completes the separation. The septum dipole is followed by a series of Panofsky quadrupole magnets to keep the horizontal dispersion under control. The downstream optics includes three vertical chicanes with the peak dispersion of 6.8 cm, 6.8 cm and 2 cm respectively. The first chicane is used for collimation of the disrupted low energy tail, and the next two are included for energy and polarization diagnostics. The quadrupoles after the energy chicane are adjusted to get a second focus at the centre of the polarization chicane. The angular transformation term R22 from the IP is adjusted to +0.5, one of the preferred values for the polarization measurement. The extraction line after the chicanes includes horizontal bends to create the required transverse separation for the beam dump from the incoming line and will include few quadrupoles to increase the undisrupted beam size on the beam dump. The linear optics of the extraction line is shown in Fig. 5. The disrupted beam was tracked up to first vertical chicane using the tracking code TURTLE. The disrupted beam phase space was generated using GUINEA-PIG for the nominal and low power parameter sets. In the simulations the separator electrodes are offset by 1 cm towards the low energy side of the bend to reduce the loss of low energy radiative Bhabhas and disrupted beam particles on the separator plates.

There are two philosophies for designing the collimation system: (1) place protection collimators, some of which must withstand several hundred kW, in front of every optical element in the entire extraction line, or (2) place a single high power dump at a location where most of the disrupted beam and beamstrahlung tails can be collimated, leaving the need for only a few lower power collimators in the rest of the extraction line. In addition, the high power beamstrahlung cone is going back through the incoming beam line and the beamstrahlung tails must be collimated in order to limit the shared magnet apertures to a reasonable size. In the present design, sufficient separation of the incoming beam and the outgoing beamstrahlung occurs at about 300 m from the IP.

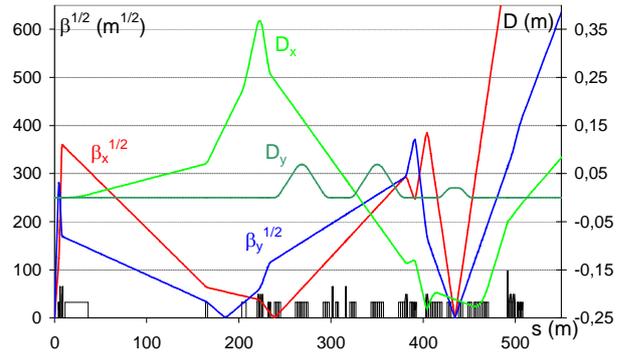

*Figure 5: Linear optics of the extraction beam line.*

As seen in Table 1, the present study uses option (2) with most of the collimation occurring at a single location, called the Intermediate Dump. This collimator, which has holes for the incoming and outgoing beams, would be modelled after a similar aluminum/water 2.2 MW device at SLAC and located at a place where it can be well shielded to protect the environment and nearby beam line elements from radiation damage. The collimator inboard of B1 and the Intermediate Dump must also be able to withstand temporary larger charged and beamstrahlung losses, given by the second set of power levels, when the beam-beam vertical deflection is at a maximum.

*Table 1: Power lost at different locations for Nominal and Low Power parameter set.*

| Loss point | Power lost (kW) @ 500 GeV CM Nominal / Low P parameter set | | |
|---|---|---|---|
| | Charged | Beam-strahlung | Bhabhas |
| QD0/SD0 | 0 / 3. 10$^{-5}$ | 0 / 0 | 13. 10$^{-6}$ |
| QF1/SF1 | 0 / 5. 10$^{-4}$ | 0 / 0 | 17. 10$^{-6}$ |
| Synch. Mask | 0 / 2. 10$^{-3}$ | 0 / 0 | 35. 10$^{-5}$ |
| E.S. plates | 0 / 0.7 | 0 / 0 | 33. 10$^{-5}$ |
| PC inboard B1 | 0.001 / 0.8 13 / 130 | 0 / 0 0.05 / 25 | 8. 10$^{-5}$ |
| Intermediate Dump | 25 / 237 101 / 397 | 57 / 105 162 / 262 | 8. 10$^{-3}$ |
| 5mrad dipole | 0 / 0 | 0 / 0.3 | 0 |
| 5mrad dipole low E side | 0.5 / 1.5 | - / - | 14. 10$^{-4}$ |
| Beamstrahlung Dump | - / - | 207 / 209 | - |
| Charged Dump | 11.275 / 11.060 | - / - | - |

## OUTLOOK

The ILC head-on collision scheme is an attractive scheme from the point of view the final focus operation and detector performance, and possibly overall cost.

Besides further optimisation, several aspects need to be studied to assess its feasibility:
- The overvall 1 TeV and high luminosity upgrades.
- The design and hardware performance of the electrostatic separator at 1 TeV and the impact of possible breakdowns on the machine protection.
- The design and hardware performance of the intermediate collimator at about 0.5 MW power.
- The performance of the diagnostics once the phase space distributions at the secondary focus have been computed.
- The effect of the large aperture B1 and B2 soft dipole fringing fields on the outgoing trajectory.
- The interplay between IP fast feedback corrections and the parasitic beam-beam kicks.